\documentclass[12pt,epsf,russian]{article}

\usepackage[dvips]{graphicx}
\usepackage{latexsym}
\usepackage{epsfig}
\usepackage{amsfonts}

\begin{document}

\begin{center}
{\bfseries  \Large Chiral String-Soliton Model for the light chiral
baryons} \vskip 5mm Oleg Pavlovsky $^\dag$ $^\ddag$, Maxim
Ulybyshev$^\dag$ \vskip 5mm {$\dag$ \small {\it Moscow State
University, Moscow, Russia} \vskip 1mm $\ddag$ \small {\it Institute
for Theoretical and Experimental Physics, Moscow, Russia} \vskip 1mm
{\it E-mail: ovp@goa.bog.msu.ru, ulybyshev@goa.bog.msu.ru} } \vskip
5mm
\end{center}

\begin{center}
\begin{minipage}{150mm}
\centerline{\bf Abstract} The Chiral String-Soliton Model is a
joining of the two notions about the light chiral baryons: the
chiral soliton models (like the Skyrme model)  and the Quark-Gluon
String models. The ChSS model is based on the  Effective Chiral
Lagrangian which was proposed in \cite{Simonov}. We have studied the
physical properties of the light chiral baryon within the framework
of this ChSS model.
\end{minipage}
\end{center}

\vskip 10mm

\section*{Introduction}

The problem of finding an appropriate theoretical model for the
light baryons is a one of the essential long-standing tasks in
hadron physics. Many models of the light baryon were proposed so
far. Ideologically all these models can be divided by
 two subclasses. The first group of models is based on the idea of
the realistic chiral solution which firstly was proposed by Skyrme
\cite{Skyrme}. The baryon is treated within this approach as a
topological soliton of the non-linear chiral meson field. The Chiral
bag model \cite{Hosaka:1996ee} which was very popular during last
time is a development of this idea. Despite of the fact that chiral
bag model can describe the quark degrees of freedom in the baryon,
there are no natural picture of the confinement in these models. The
simulation on the lattice explicitly shows us that the confinement
phenomenon in QCD is connected with the formation of the gluon
string between the quarks. This phenomenon can be described
phenomenologically by means of the effective linear interaction
between the quarks. The models of such type form the second class of
the models of the light baryons. A weak place of such models is that
it is difficult to take into the account the chiral degrees of
freedom of the meson cloud around baryon. It is very sad because the
chiral effects play the essential role in the physics of light
baryons. So the question is how to combine these two different
points of view on the light baryons into the framework of a single
model?

An important step has been made recently towards such model
\cite{Simonov}. The effective chiral Lagrangian which contains the
effects of both confinement and chiral symmetry breaking was
proposed. It was shown that this Lagrangian is reproduced the
Gell-Mann-Oakes-Renner relations for the light mesons and correctly
describes the spectrum of radial excitations of $\pi$ and $K$
mesons. In our work we use this model for description of the light
baryons.

\section{Effective Chiral Lagrangian and soliton model}
In this paper we explore bound states of valence quark using the
chiral invariant effective $\sigma$-model lagrangian. Our model is a
rather similar to the Soliton-Bag model \cite{Lee:1991ax}. We allow
the additional string potential for quarks to realize the
confinement phenomenon. The structure of our effective lagrangian is
connected with the effective chiral theory which was derived from
QCD in the framework of the field correlator method. Using the this
method one can integrate out gluon field $A_{\mu}$ and find the
contributions from fermion determinant which generate the meson
effective lagrangian. As a result one can get
$$
{\cal L}={\cal L}_\psi + {\cal L}_\pi + {\cal L}_{{int}},
$$
where ${\cal L}_\psi = \overline{\psi}(i \gamma
\partial-m_0)\psi$ and ${\cal L}_\pi $ is a self-interacting part of the meson
field. The bullet point of the model is the interaction part of
the lagrangian density ${\cal L}_{{int}}$:
$$
{\cal L}_{{int}}= M_{eff}(\vec{x},\vec{y})\exp (i \gamma_5
(\vec{\tau} \vec{\varphi})),
$$
where $M_{eff}$ is the effective quark mass operator. It was shown
in \cite{Simonov} that
$$
M_{eff} \approx \sigma \mid \vec{x} - \vec{y} \mid
$$
for large distance between quarks; $\sigma$ is the string tension.
The physical interpretation of this result is quite clear. If the
distance between the quarks becomes large, the effective quark-meson
coupling becomes large too. This effect is not unexpected by reason
of the string broken phenomenon with the production of the light
meson.

Now we promote this idea of the effective quark-meson coupling for
the baryon physics. In the first approximation let us consider the
quarks which live in the self-consistent linear potential with the
center in the point of the string junction. It means that in this
approximation we neglect the contributions from quark-antiquark
interaction in comparison with the interaction with the string
junction. It is so-called {\bf Y}-type picture of the baryon.

Let us consider the self-interacting, spin-0, isosinglet field
$\phi(x)$, and the isotriplet pion field $\vec{\pi} (x)$ which
interacts with the isodouplet, spin-$1 \over 2$ quark field
$\psi(x)$. The Lagrangian density ${\cal L}$ can be written as
$$
{\cal L}={1 \over 2} \partial_\mu \phi \partial^\mu \phi + {1
\over 2} \partial_\mu \vec{\pi} \partial^\mu \vec{\pi} +
\bar{\psi} (i \gamma_\mu \partial^\mu - m_0)\psi - M_{eff}
\bar{\psi} [(\phi + i \gamma_5 (\vec{\pi} \vec{\tau}))/f_\pi ]
\psi - {\lambda^2 \over 4} (\phi^2+\vec{\pi}^2-f_\pi^2)^2.
$$
For the light quarks $m_0 \ll \sqrt{\sigma}$ so let us take $m_0=0$.

In the main-field approximation the $\phi$ and $\vec{\pi}$ field
are taken as classical, time-independent $c$-number  fields. The
quark field is expanded in single-particle modes $\psi_n$ which
satisfy the Dirac equation
\begin{equation}
 -i \vec{\gamma} \vec{\partial}\psi_n -
M_{eff} [(\phi + i \gamma_5 (\vec{\pi} \vec{\tau}))/f_\pi ]
\psi_n= {\cal E}_n \psi_n. \label{Dirac}
\end{equation}

If one put $N$ quarks into the lowest mode with energy ${\cal
E}_0$, the total energy of the system is given by
\begin{equation}
E_{tot}=N {\cal E}_0 + \int d^3 x [  {1 \over 2} (\vec{\partial}
\phi)^2  + {1 \over 2} (\vec{\partial}\vec{\pi})^2 + {\lambda^2
\over 4} (\phi^2+\vec{\pi}^2-f_\pi^2)^2 ]. \label{Tot_enegry1}
\end{equation}
The total energy $E_{tot}$ is a functional of the chiral fields
$\phi$ and $\vec{\pi}$. At equilibrium point, the energy is
stationary with respect to variations of the fields. One can get the
equations of motion for meson fields by the minimizing of $E_{tot}$
\begin{equation}
-\vec{\partial}^2 \phi+\lambda^2 (\phi^2+\vec{\pi}^2-f_\pi^2)\phi
= - N M_{eff}/f_\pi (\bar{\psi}_0 {\psi}_0) \label{eq_phi}
\end{equation}
\begin{equation}
-\vec{\partial}^2 \vec{\pi}+\lambda^2
(\phi^2+\vec{\pi}^2-f_\pi^2)\vec{\pi} = - N M_{eff}/f_\pi
(\bar{\psi}_0 i\gamma_5 \vec{\tau} {\psi}_0). \label{eq_pi}
\end{equation}

\section{The hedgehog solution and the chiral angle parametrization}

The equations (\ref{Dirac}),(\ref{eq_phi}) and (\ref{eq_pi}) are the
self-consistent set of equations for Dirac orbitals $\psi_n$ and
chiral fields $\phi$ and $\vec{\pi}$. In order to obtain the minimum
energy solution, let us consider the spherically symmetrical
hedgehog ansatz for the chiral fields
$$
\phi = \phi(r), \, \, \vec{\pi} = \vec{r} \pi(r)
$$
and the $s$-state solution for Dirac orbital
$$
\psi= {{u(r)} \choose {i(\vec{\sigma} \vec{r})v(r)}} \chi
$$
where $\chi$ is a state in which the spin and isospin of the fermion
couple to zero. After substituting this ansatz into the equation of
motion one can get:
$$
- {1 \over r} {d^2 \over dr^2 } (r \phi) + \lambda^2
(\phi^2+\pi^2-f_\pi^2) \phi = - N{1 \over f_\pi} \sigma r
(u^2-v^2)
$$

$$
 - {1 \over r} {d^2 \over dr^2 } (r \pi) +  {2 \over r^2} \pi +
\lambda^2 (\phi^2+\pi^2-f_\pi^2) \pi = - 2N{1 \over f_\pi} \sigma
r uv
$$

$$
{d u \over dr}  +   \left({\cal E}+{1 \over f_\pi}\sigma r \phi
\right) v + {1 \over f_\pi}\sigma r \pi u  = 0
$$

$$
{d v \over dr}  +  {2 \over r} v + \left( -{\cal E}+{1 \over
f_\pi}\sigma r \phi \right)u -  {1 \over f_\pi}\sigma r \pi v  = 0
$$

If the constant $\lambda$ is large enough, the chiral fields
restrict to the chiral circle $ \phi^2+\pi^2=f_\pi^2 $. In this
case it is possible to parameterize the chiral fields by means of
a chiral angle $\theta(r)$:
$$
\phi (r) = f_\pi \cos \theta (r), \, \, \, \, \, \, \, \, \pi(r) =
f_\pi \sin \theta (r).
$$
The equations of motion in terms of the chiral angle are:
\begin{equation}
\theta'' +  {2 \over r} \theta' -  {1 \over r^2} \sin 2\theta  = -
{N \over f_\pi^2} \sigma r [(u^2-v^2)\sin\theta+2uv\cos\theta ]
\label{eq_teta_1}
\end{equation}
\begin{equation}
{d u \over dr}  +   \left({\cal E}+\sigma r \cos\theta \right) v +
\sigma r \sin\theta u  = 0 \label{eq_teta_2}
\end{equation}
\begin{equation}
{d v \over dr}  +  {2 \over r} v + \left( -{\cal E}+\sigma r
\cos\theta \right)u -  \sigma r \sin v  = 0 \label{eq_teta_3}
\end{equation}
The fermion wave function is normalized to
$$
4 \pi \int^\infty_0 r^2 dr (u^2+v^2)=1
$$

The energy of the system in terms of the chiral angle is:
\begin{equation}
E_{tot}=N{\cal E}+ 2 \pi f_\pi^2 \int^\infty_0 r^2 dr (\theta'^2 +
{2 \over r^2}\sin^2\theta) \label{energy_teta}
\end{equation}

The chiral angle approximation is a very interesting due to
numerical and theoretical reasons. We can consider this
approximation as a first step of our consideration.

\section{Chiral String-Soliton Model: numerical results}
Before we start to discuss the numerical results of our simulation
of the equations of motion (\ref{eq_teta_1}), (\ref{eq_teta_2}) and
(\ref{eq_teta_3}) let us consider the Dirac particle in the string
potential without chiral soliton. One obtains the equations of
motion in this case by substituting the vacuum value $\theta=0$ into
(\ref{eq_teta_2}) and (\ref{eq_teta_3}):

\begin{equation}
{d u \over dr}  +   \left({\cal E}+\sigma r  \right) v  = 0
\label{eq_pure_1}
\end{equation}
\begin{equation}
{d v \over dr}  +  {2 \over r} v + \left( -{\cal E}+\sigma r
\right)u = 0 \label{eq_pure_3}
\end{equation}

The energy spectrum in this case can be found numerically and the
lowest energy level is equal to ${\cal E}_0= 1.61 \sqrt{\sigma}$.
The value of $\sqrt{\sigma}$ was well studied in the lattice
calculations and varies from 400 MeV to 420 MeV. Thereby the total
energy of the system for $N=3$ is
$$
E_{str}= N {\cal E}_0 \approx \mathrm{2} \, GeV,
$$
and this is obviously much more than the physical scale of the light
baryon mass (about 1 GeV). Of course this fact is a sequence of the
ignoring of the chiral effects. And the main motivation of our work
is to show how such effects can be taken into account. Now we will
demonstrate  that the chiral soliton around baryon leads to
diminishing  of the total mass of the system down to the physical
value about 1 GeV.

Now let us consider the set of equations  (\ref{eq_teta_1}),
(\ref{eq_teta_2}) and (\ref{eq_teta_3}). As we point out the soliton
sector for chiral field will be interesting for us. The topological
chiral soliton corresponds to the following boundary condition:
$$
\theta(r) \to \pi, \, \, \, \mbox{as} \, \, \, r \to 0;
$$
$$
\theta(r) \to 0, \, \, \, \mbox{as} \, \, \, r \to \infty.
$$

\begin{figure}[h]
 \begin{center} \epsfbox{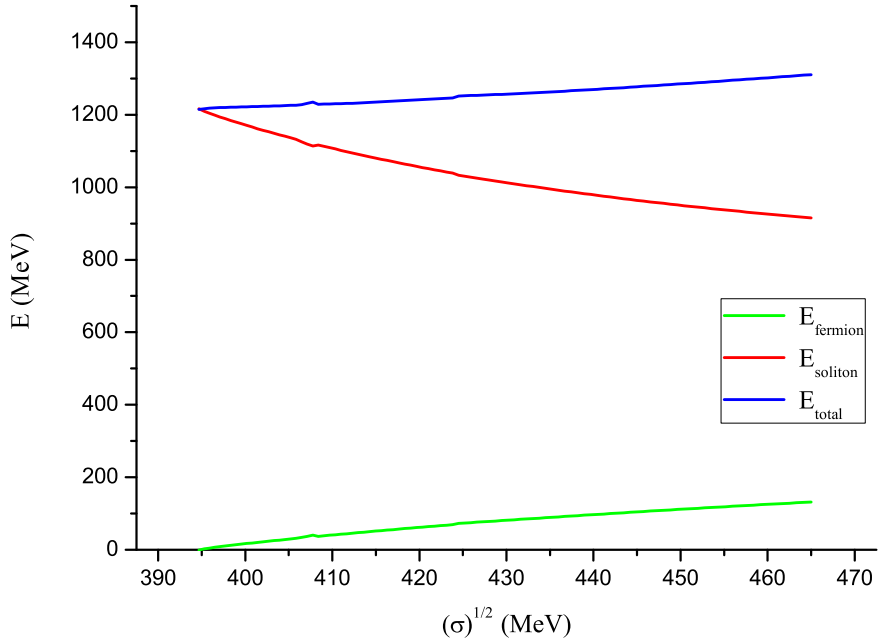}
 \end{center}
 \caption{The total energy of the system, the energy of the chiral soliton and the energy
 of the Dirac orbital as functions of $\sqrt{\sigma}$}
\end{figure}

The numerical results for the total energy of the self-consistent
system of the $N=3$ massless fermion and chiral topological soliton
as the function of $\sqrt{\sigma}$ are presented in Fig. 1.

First of all, we see that for physical value of $\sqrt{\sigma}$ the
total mass of the baryon is about 1230 MeV. It is close to the
experimental data for the mass of $\Delta$-baryon: $M_\Delta=1232$
MeV. Of course, is such simple model there is no possible to
reproduce mass of the proton directly because the spin effects are
not taken into account. It is possible to do by using of the
standard technique \cite{Witten} and it will be the object of our
next studies.

Another interesting feature of this result is that for the physical
values of the string tension $\sigma$ the energy of the Dirac
orbital ${\cal E}$ is very close to zero (${\cal E}=0$ for
$\sqrt{\sigma}=394$ MeV). It means that the mass of the light
baryons in this model almost forms due to the chiral soliton. The
Dirac orbital is very essential for the stability of the system but
it is clear that the meson cloud plays the dominant role in the
dynamics. This fact is very interesting and intriguing, first of all
due to the possible connection with the famous "proton spin"
paradox. We plan to discuss it in the future.

\begin{figure}[t]
 \begin{center} \epsfbox{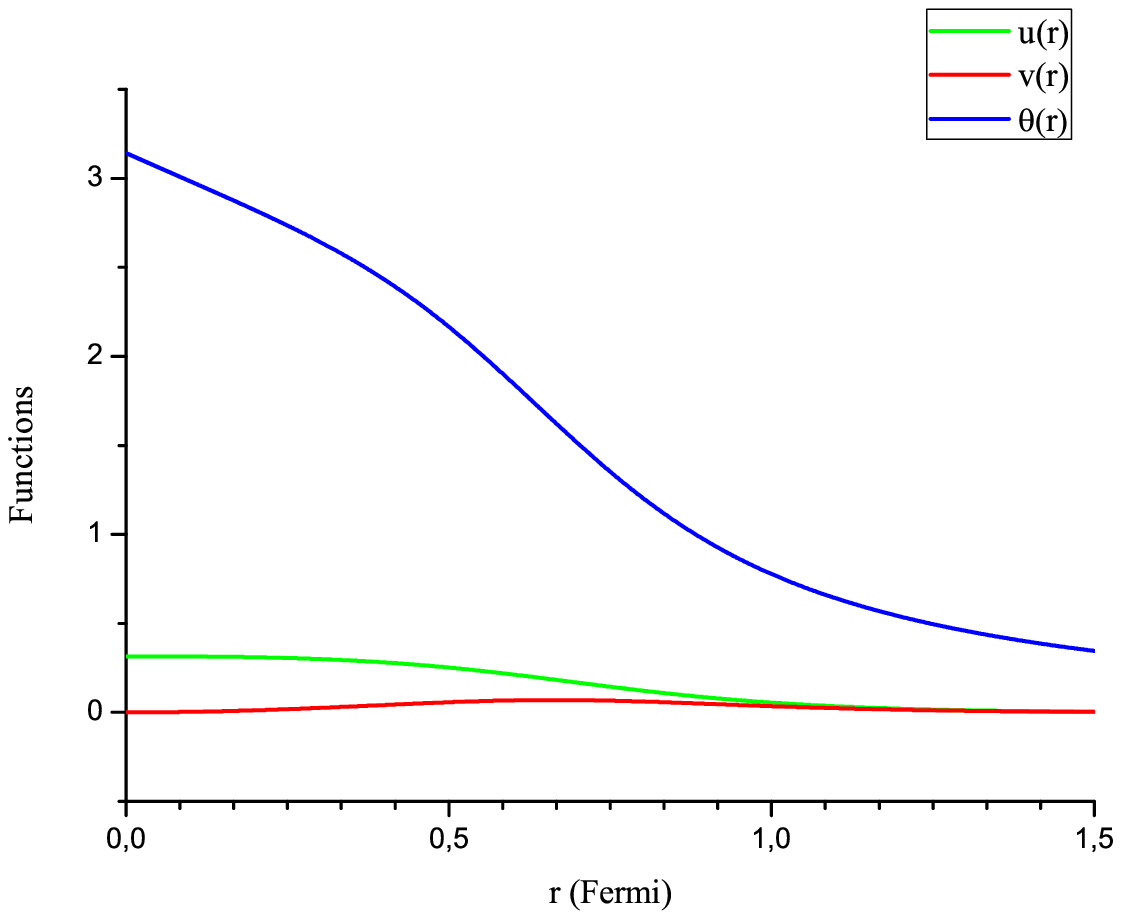}
 \end{center}
 \caption{The radial functions $u(r)$, $v(r)$ and $\theta(r)$}
\end{figure}

\begin{figure}[t]
 \begin{center} \epsfbox{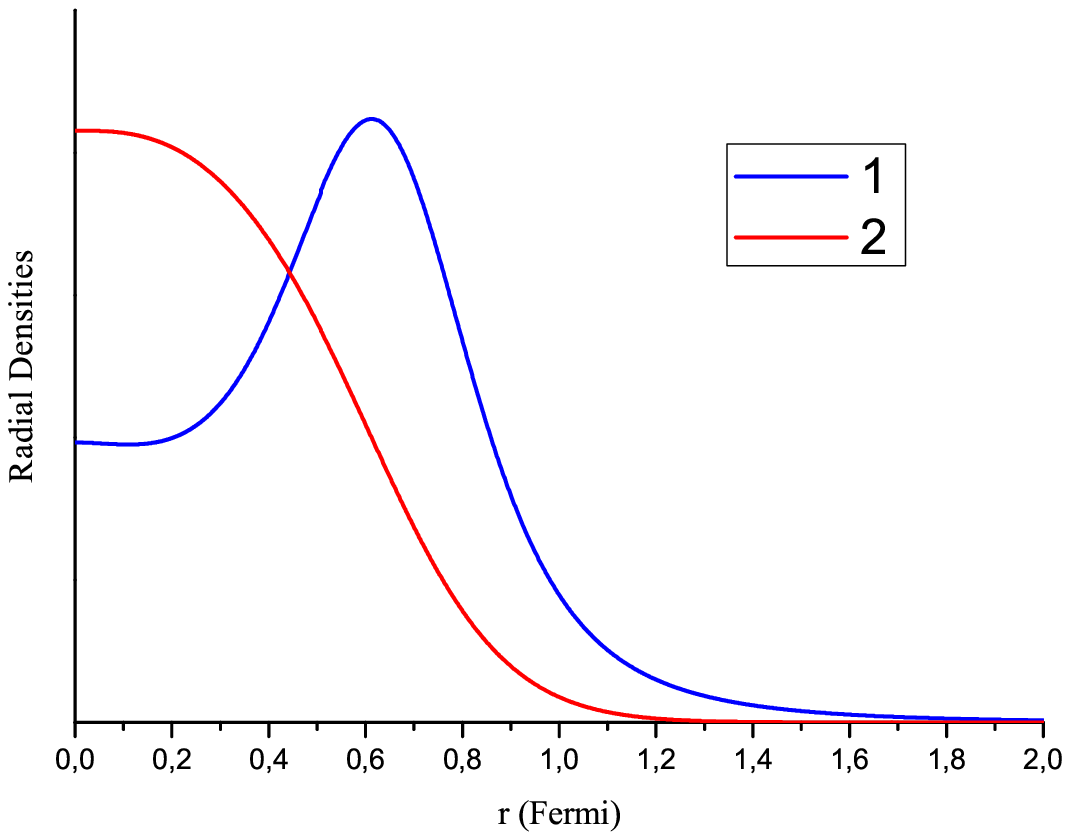}
 \end{center}
 \caption{The radial density of the meson (1) and fermion (2) fields: $(1)= (\theta'^2 +
{2 \over r^2}\sin^2\theta)$ and $(2)=u^2+v^2$}
\end{figure}

In Fig. 2 and Fig. 3 we show the radial functions and densities
for the bound state of zero mode (${\cal E}=0$ for
$\sqrt{\sigma}=394$ MeV). We see that the fermions are localized
in a small region in the center and meson cloud surrounds this
quark kernel.

\section*{Conclusion}
In this paper we have  studied the properties of the light baryon in
the framework of the Chiral String-Soliton Model. In main-field
approximation we interpret  the baryon as the self-consistent bound
state of the three valent quarks and meson soliton. We have shown
that the generation of the meson cloud is energetically profitable
in comparison with the bound state without chiral soliton. The
Chiral String-Soliton Model leads to the good prediction for the
mass of the light baryon.

Finally, we would like to emphasize that the ChSS model has no free
parameter. Only the string tension one can vary in very short
region. It means that the prediction power of this model is very
strong. Another advantage of this approach is that the basis of this
model is Effective Chiral Lagrangian which can derived from QCD
\cite{Simonov} and works very well in the meson sector. In our work
we show that this Lagrangian is applicable for the physics of the
light chiral baryon too.


\begin{thebibliography}{99}

\bibitem{Simonov}
  S.~M.~Fedorov and Yu.~A.~Simonov,
  JETP Lett.\  {\bf 78}, 57 (2003)
  [Pisma Zh.\ Eksp.\ Teor.\ Fiz.\  {\bf 78}, 67 (2003)]
  [arXiv:hep-ph/0306216].

  \bibitem{Skyrme}
  T.~H.~R.~Skyrme,
  Nucl.\ Phys.\  {\bf 31} (1962) 556.

\bibitem{Hosaka:1996ee}
  A.~Hosaka and H.~Toki,
  Phys.\ Rept.\  {\bf 277} (1996) 65.

  \bibitem{Lee:1991ax}
  T.~D.~Lee and Y.~Pang,
  Phys.\ Rept.\  {\bf 221} (1992) 251.

\bibitem{Witten}
  G.~S.~Adkins, C.~R.~Nappi and E.~Witten,
  Nucl.\ Phys.\  B {\bf 228} (1983) 552.


\end{thebibliography}
\end{document}